\documentclass[twocolumn, tighten, astrosym, twocolappendix]{aastex631}

\usepackage{amsmath}
\usepackage{amssymb}
\usepackage{graphicx}

\usepackage{array, tabularx}
\usepackage{mathscinet}

\usepackage{savesym}
\savesymbol{tablenum}
\usepackage{siunitx}
\usepackage{xspace}

\newcommand{\THK}{\textit{Ts\rasp\hspace{-2pt}eit HaKokhavim}\xspace}
\newcommand{\MS}{\textit{Motsa\rasp\hspace{-2pt}ei Shabbat}\xspace}

\restoresymbol{SIX}{tablenum}

\begin{document}

\title{The Astronomy of Halakhic Nightfall: \\ Calculating \THK and \MS}

\shorttitle{Timing of Nightfall}
\shortauthors{Aster G. Taylor}

\author[0000-0002-0140-4475]{Aster G. Taylor}
\correspondingauthor{A. G. Taylor}
\altaffiliation{Fannie and John Hertz Foundation Fellow}
\affiliation{Dept. of Astronomy, University of Michigan, Ann Arbor, MI 48109}
\email{agtaylor@umich.edu}

\begin{abstract}

In Jewish tradition, the boundary between days is not midnight, but nightfall. Nightfall is when one must pray \textit{ma\lasp\hspace{-2pt}ariv}, count the omer, and may engage in activity after Shabbat. While there are slightly different definitions of nightfall for Shabbat restrictions (\MS) versus other purposes (\THK), most minhagim define nightfall to be when three stars of some specified size {and proximity} are visible in the sky. This definition presents some difficulty. Not only are these conditions difficult to {define}, they are further complicated by weather, cloud cover, and light pollution. Observant Jews therefore usually approximate these times by defining nightfall to be when the sun has reached a certain {distance} below the horizon. Although these approximations have been effective, modern astronomy enables us to {calculate precisely when these conditions are met}. This work applies these techniques to calculate the time of \THK and \MS, {evaluates the accuracy of the approximations, and explores the effects of light pollution on these times.} {While \THK is reasonably well-approximated by standard methods, the conditions for \MS generally occur after the time predicted by the approximations.} Light pollution does not shift \THK by more than a minute but significantly changes \MS. In fact, light pollution causes the conditions of \MS to never be met on at least some nights in every population center. Finally, I provide a tool (\href{halakhic-nightfall.streamlit.app}{halakhic-nightfall.streamlit.app}) for calculating {\THK and \MS} at an arbitrary {time and} location {on Earth}.
\end{abstract}


\section{Introduction}

Due to its {origins in antiquity}, the rituals and restrictions of Jewish practice are defined in terms of astronomical observations. Of particular importance is nightfall (\THK), which separates one day from the next. The nightly repetition of the Shema prayer, the counting of the omer during the Seven Weeks, the end of the fast days, and the relaxation of the strictures of Shabbat and other \textit{yamim tovim} occur at nightfall. Due to the religious significance of this time, its determination is of critical importance. 

However, the {halakhic} definition of nightfall is somewhat uncertain. The Talmud recognizes an intervening period between sunset (\textit{shqi\hspace{1pt}\lasp\hspace{-2pt}ah}, when the day ends) and nightfall, when the night begins. \textit{Bein HaShmashot}, the twilight period, is of uncertain religious status. While this issue is {halakhically} resolved by applying the stringencies of both days to the twilight period (b. Shabbat 34b), there is no unambiguous astronomical event that divides the twilight from the night in the same way that sunset divides the day from the twilight.

Three definitions for nightfall are given in the Talmud --- four \textit{mil} after sunset (b. Pesahim 94a), three-quarters of a \textit{mil} after sunset (b. Shabbat 34b), or when three ``medium'' stars are visible (\THK, b. Shabbat 35b).\footnote{
Note that a \textit{mil} is a unit of distance, and it is assumed that when a time frame is given in \textit{mil}, it means the time it would take a person to walk that distance. Generally, one \textit{mil} is expected to be \qty{18}{\minute} (Shulchan Aruch Orach Chaim 459:2).}
There are generally three approaches to resolving these discrepancies. While this introduction ignores several trains of rabbinic thought and minhagim, the discussion presented below captures the broad overview.

Rabbeinu Tam holds that there are two sunsets. The discrepancies in the length of twilight thus refer to the times between {each of these} sunsets and nightfall. In R. Tam's approach, nighttime begins four \textit{mil} after astronomical sunset for almost all purposes, and the emergence of three stars is subject to too much interpretation to be reliable. While there is further debate on how to understand the time of four \textit{mil} in R. Tam's interpretation, the general standard is to assume a fixed \qty{72}{\min} delay between sunset and the onset of nightfall. Note that Rabbi Moshe Feinstein has ruled that the \qty{72}{\min} rule only applies in Europe, and that in New York all the stars are visible by \qty{50}{\min} after sunset (Shu''t Igrot Moshe Orach Chaim 4:62). The Vilna Gaon (the Gra) makes a similar interpretation {to that of R. Tam}, but takes the general nightfall time to be three-fourths of a \textit{mil} after the sunset (Biur HaGra, Orach Chaim 261:1-2), with a variable \textit{mil} time. While variants of R. Tam's opinion are followed by a minority of communities, the Gra's opinion is rarely observed.

The final approach is that of Rav Yeḥi\rasp\hspace{-1.5pt}el Michael Tuqatsinsqi,\footnote{More common transliterations include Tukachinsky, Tuktsinski, and Tucazinsky.} who notes that the opinions of R. Tam and the Gra are against almost all of the text of the Gemara. As a result, R. Tuqatsinsqi holds that nightfall begins with the emergence of three medium stars. However, the difficulties of observing stars and {the fact that a ``medium'' star is not well-defined} makes this methodology practically complex. In addition, observations of stars are complicated by weather, buildings, and (in the modern world) light pollution. Given these complexities, R. Tuqatsinsqi advocates for an expert to determine the timing of when stars are visible in a given location (Bein HaShemashot 2:8 p. 29). This measurement can then be extended to cover an arbitrary location on Earth by asserting that the visibility of stars is entirely determined by the altitude of the Sun. If an expert observes \THK when the Sun is at an altitude $h$ below the horizon at their location, communities around the world then assume that \THK occurs when the Sun is at $h$ across the world.

An additional complexity is introduced by the Shulchan Aruch, which {adds that} on Shabbat
\begin{quote}
    [o]ne must be careful not to do work until he sees three small stars that are not scattered, but rather in a row in one place (Shulchan Aruch Orach Chaim 293:2).
\end{quote}
The time when three small stars are visible ``in a row in one place'' (\MS) is therefore distinct from the time of \THK, when three medium stars are visible at any point in the sky. 
The phrase translated here as ``in a row in one place'' is \textit{mfuzar ela ratsuf}, which can also be rendered as ``scattered but continuous''. 
With this framing, this condition can be understood to mean that the stars should be somewhat close together, but not necessarily in a straight line. 
Indeed, there are vanishingly few sets of three nearby colinear stars, and almost all sets of three that are somewhat close together will appear to be ``scattered but continuous''. 
Finally, texts of the halakhah generally reformulate this phrase to require ``clustered'' or ``grouped'' stars (e.g., Peninei Halakhah, Shabbat 3:2:6), in line with this interpretation. 
The Mishna Brurah further explains that the more stringent definition {of} \MS is imposed because nobody is currently enough of an expert to determine the time of \THK. 
Therefore, one should be careful and avoid proscribed actions until \MS (Mishna Brurah 293:2). 

Most Jewish communities follow R. Tuqatsinsqi and use separate solar altitude approximations for \THK and \MS. 
In general, \MS is assumed to occur when the Sun is \qty{8.5}{\degree} below the horizon. 
{Note that ``altitude'' here refers to the number of degrees between an object in the sky and the horizon, while ``elevation'' will be used to refer to the height of a location above sea level.}
By a similar calculation, the time of \THK is generally assumed to occur when the Sun is \qty{6.45}{\degree} below the horizon.\footnote{
See, e.g., \href{https://www.myzmanim.com/search.aspx}{myzmanim.com} and the \href{https://kosherjava.com/}{KosherJava} plugin, which present the time of \THK using the \qty{8.5}{\degree} calculation. 
R. Dovid Eisikowitz (\href{https://www.myzmanim.com/search.aspx}{myzmanim.com}) has shown that R. Moshe's \qty{50}{min} limit is consistent with the \qty{8.5}{\degree} definition.}

There are several factors that complicate this approximation. 
First, the sky brightness during twilight is not necessarily identical across locations, even if the Sun is at the same altitude below the horizon. 
In particular, the elevation of the site and the density of the atmosphere can significantly modify the relative brightness of the eastern and western skies. 
Second, the advent of electric lighting has made light pollution commonplace and significantly modified the night sky brightness. 
Today, only about $20\%$ of the global population lives under an unpolluted night sky \citep{Falchi2016a}. 
Finally, the small number of {bright} stars {that meet the relevant conditions} means that the nightfall times can vary even for identical solar altitudes.

While there have been extensive previous studies on celestial visibility and the effects of light pollution (e.g., \citealt{Garstang1986, Schaefer1990, Schaefer1992}, see \citealt{Rozenberg1966} or \citealt{Barentine2022} for a review) and there has been discussion of the impact of celestial visibility on religious calendars (see the discussion of Islamic calendars and the timing of the Crucifixion in \citealt{Schaefer1993}), {as far as the author is aware} the modern science of celestial visibility has not been applied {in a} Jewish context. 
In this paper, I introduce a tool\footnote{Available on the web at \href{halakhic-nightfall.streamlit.app}{halakhic-nightfall.streamlit.app}.} that calculates when three stars of the relevant size and proximity are visible at a given location and date. This package is then used to compare the timing of \THK and \MS to the solar altitude approximations and to determine the significance of light pollution.

This paper is structured as follows: The definitions of star size and ``in one place'' are specified and justified in Sec. \ref{sec:condspec}. Sec. \ref{sec:skybri} then calculates the brightness of the twilight sky and the visibility of stars in a given sky and describes the algorithm used to calculate the nightfall timing. {In Sec. \ref{sec:tcomp},} the results are presented and compared to the approximations. The effects of light pollution are specified in Sec. \ref{sec:lp}. Sec. \ref{sec:pareffs} discusses the effects of varying the {definitions of ``medium'', ``small'', and ``in one place''} on the timing of \THK and \MS. Finally, Sec. \ref{sec:conc} concludes with a summary and a discussion of these results. 

\section{Nightfall Condition Specification}\label{sec:condspec}

In order to begin, some definitions must be made. The calculation of nightfall relies on both a definition of star sizes and ``in one place'', both of which are rather vague. 
The size of a star will be defined in terms of its apparent magnitude\footnote{
Referring to the brightness of a star as viewed from the Earth. This is in contrast to the absolute magnitude, which describes the actual brightness of a star independent of its distance from the Earth.}
and the ``in one place'' condition will be defined in terms of {the angular} distance {between stars} in the sky. 
Given the technical nature of these definitions, only an expert can determine which stars fall into which size category by observation. 
Using the apparent magnitude is thus consistent with R. Tuqatsinsqi's implication that the separation between small and medium stars is something that only an expert can reliably know. 

The apparent magnitude of a star is a single number between $-\infty$ and $\infty$ that describes the brightness of the star as viewed from the Earth, based on the original star catalog of Hipparchus (c. 190 - c. 120 BC). 
While this system {is currently understood to} specify the brightness of stars, it was originally used to specify the apparent width {or} size of stars, {since brighter stars will look larger to the eye}. 
Somewhat unintuitively, the apparent magnitude is a negative logarithmic scale. 
To demonstrate, consider two stars A and B, where star A has a magnitude that is $1$ smaller than star B. 
Then {by definition} the Earth receives $2.5$ times more energy from star A than star B, meaning star A looks brighter.

This scale is chosen to closely match the human perception of brightness, which scales logarithmically with the amount of light received.
Aside from the Sun, there are 5 stars with an apparent magnitude less than \qty{0}{mag}, and the dimmest stars visible in a pristine night sky have a magnitude of $\sim\qty{6}{mag}$. 

When discussing ``medium'' stars, the Talmud specifies that 
\begin{quote}
    [t]his is neither referring to large stars that are visible even during the day, nor to small stars that are visible only late at night. Rather, it is referring to medium-sized stars (b. Shabbat 35b).
\end{quote}
A medium star is thus defined to be a star that {does not meet the criteria for a large or small star}. 
Since the day ends at sunset (see b. Shabbat 34b), the boundary between a ``large'' and a ``medium'' star is the dimmest star that can be seen at sunset.

This cutoff depends on whether the observer knows the location of all of the large stars or is blindly searching.
If the observer knows the location of these stars, they can use averted vision, a technique where a trained observer looks slightly to the side of the target object. 
Since the rods on the outer edge of the retina are more sensitive to small differences in illuminance, this technique allows for the detection of dimmer objects than directed vision would allow. 
However, averted vision requires that the observer know the precise location of the target object and can easily focus on it.
In the absence of the Moon and light pollution, the minimum (brightest) limiting magnitude is $\sim\qty{2}{mag}$ with averted vision. 

In the author's opinion, it is not unreasonable to expect the expert observer advocated by R. Tuqatsinsqi to know the locations of the \num{50} stars brighter than \qty{2}{mag} and therefore be able to see these stars even at sunset {using averted vision}. 
Based on these considerations, all stars with magnitudes $m\leq\qty{2}{mag}$ will be be ``large'' for the purposes of this work, while those with $m>\qty{2}{mag}$ are either ``medium'' or ``small''. 

{When looking for small stars, }the observer is more likely scanning the sky for stars without a known target. 
{After all, it is difficult for even an expert to know the locations of the thousands of dim stars in the sky and look for them directly.}
In such a circumstance, some additional contrast is necessary for a star to be noticeable. 
The limiting noticeable magnitude {for small stars} is therefore taken to be one magnitude lower (brighter) than the value given by Eq. \eqref{eq:maglim}. This additional factor of 2.5 in brightness is consistent with common experience and previous reports on stellar observability (see, e.g., \citealt{Tousey1948, Tousey1953, Morison2017}). 

Now, additional information is necessary to provide a workable definition of a ``small'' star.
Since nightfall is defined to occur when medium stars are visible, the dimmer small stars must by definition only be visible after nightfall.
Therefore, ``late at night'' for the purposes of defining small stars must come from a separate definition. Fortunately, the Talmud adds that
\begin{quote} 
    [i]f the upper segment [of the Eastern sky] has lost its color, and its color equals that of the lower one, it is night (b. Shabbat 34b).
\end{quote}
Using this definition, the boundary between a small and medium star is the brightest star that cannot be observed when the colors of the Eastern sky near the zenith and the horizon are still different. 
Previous work on the color of the sky at twilight (\citealt{Haber2005, Zagury2012} and especially \citealt{Adams1974, Nawar1983a}) shows that the Eastern sky approaches a uniform color when the Sun is approximately \qty{8}{\degree} below the horizon. At this solar altitude, a \qty{3}{mag} star is the brightest star that cannot be seen with directed vision (once again absent the Moon and light pollution). Therefore, medium stars are those with $\qty{2}{mag}<m\leq\qty{3}{mag}$ (used for \THK) and small stars have $m>\qty{3}{mag}$ (used for \MS). {To reiterate, the cutoff for large versus medium stars assumes the observer knows the locations of the large stars and is using averted vision. On the other hand, the cutoff between medium and small stars assumes that the observer is searching blindly and requires some additional brightness to be noticed. }

The final condition needed is that the three stars are ``in one place'', which is imposed by the Shulchan Aruch (Orach Chaim 293:2) for \MS. Here, ``in one place'' is taken to mean that the three stars are each within \qty{10}{\degree} of one another.\footnote{This is approximately the size of a closed fist held at arm's length, or \num{20} full moons.} This value is chosen to approximate the foveal field of view, where vision is most precise \citep{Curcio1990}. Three stars within this distance will be visible to the observer without moving their eyes and is a reasonable definition of ``in one place''. For the sake of completeness, I will also present the nightfall time using different definitions of these parameters. 

\section{Sky Brightness and Visibility}\label{sec:skybri}

During the course of twilight on a clear night, the visibility of stars depends almost entirely on the brightness of the background sky, which sets the minimum contrast detectable by a human eye. The sky brightness during this period can be written as 
\begin{equation}\label{eq:Btot}
    B_{\rm sky}=B_{\rm night}+B_{\rm twi}+B_{\rm glare}+B_{\rm Moon}+B_{\rm lp} \,.
\end{equation}
In this equation, $B_{\rm night}=\qty{1.7e-4}{cd/m^2}$ is the sky brightness of the dark night sky \citep{Garstang1989}, $B_{\rm twi}$ is the contribution of sunlight scattering in the atmosphere, $B_{\rm glare}$ is the glare of bright point sources ({in this case,} the Moon), $B_{\rm Moon}$ is the sky brightness contribution of the Moon, and $B_{\rm lp}$ is the contribution of light pollution. {Glare is the additional brightness immediately around a bright source (such as the Moon), while the Moon itself also generally raises the sky brightness due to scattering. These two components are considered separately. }

The dominant component of the twilight sky brightness is scattered photons from the Sun. Due to the importance of multiple scattering \citep{Belikov1996}, calculating the sky brightness at twilight from first principles is difficult. Previous sky brightness calculators (e.g., \citealt{Schaefer1998, Sugerman2000}) have used an empirical equation (based on \citealt{Kastner1976}) that has been optimized to represent parameters of interest for heliacal rising (that is, looking towards the Western horizon while the Sun's altitude is between \qtyrange{0}{6}{\degree} \citealt{Schaefer1987}). However, extrapolating this formula to the remainder of {the} sky produces gross inaccuracies. This calculation therefore interpolates between empirical data \citep{Koomen1952} that provide a significantly better match for observations across the entire sky (see Appendix \ref{sec:twibrimeas}). 

These data were reported at two different sites --- Maryland (\qty{30}{m} elevation) and Sacramento Peak, New Mexico (\qty{2800}{m} elevation). Since the atmosphere's density --- and therefore scattering properties --- vary steeply with elevation, there are qualitative differences between the sky brightness measured at these locations. To account for this effect, the elevation dependence is modeled as (see \citealt{Schaefer1993} Eq. 15)
\begin{equation}\label{eq:Btwicorr}
    B(H, \vec{\theta})=B_0(\vec{\theta})\left(1-10^{-0.4k_V(H)X(H,\vec{\theta})}\right)\,.
\end{equation}
In Eq. \eqref{eq:Btwicorr}, $\vec{\theta}$ is the sky location, $H$ is the elevation, $k_V$ is the atmospheric extinction {(change in magnitude from the atmospheric absorption)} in magnitudes per airmass, and $X$ is the airmass. 
On the basis of Eq. \eqref{eq:Btwicorr}, the data for both measured sites is corrected for the elevation and averaged to find $B_0(\vec{\theta})$.
$B_{\rm twi}$ is {then} calculated with the necessary elevation correction. This function provides a good fit for prior measurements (see, e.g., \citealt{Rozenberg1966, Patat2006, Nawar2020}) and for novel measurements taken at several sites (see Appendix \ref{sec:twibrimeas}). 

For a zenith distance $Z$ and an elevation $H$, the airmass (capturing the amount of atmosphere along the line of sight) is given by \citep{Rozenberg1966, Schaefer1993}
\begin{equation}
    X(Z,H)=\left[1-\left(\frac{\sin Z}{1+(H/R_\earth)}\right)^2\right]^{-0.5}\,,
\end{equation}
where $R_\earth$ is the radius of the Earth. The atmospheric extinction $k_V$ results from a myriad of atmospheric components, but is dominated by Rayleigh scattering and atmospheric aerosols. The Rayleigh scattering component at visual wavelengths (in magnitudes per airmass) is given by \citep{Hayes1975, Schaefer1993}
\begin{equation}
    k_R=0.1066\exp\left(-\frac{H}{\qty{8.2}{km}}\right)\,,
\end{equation}
while the atmospheric aerosol component is given by \citep[to first order,][]{Krisciunas1990, Schaefer1993}
\begin{equation}\label{eq:ka}
    k_a=0.12\exp\left(-\frac{H}{\qty{1.5}{km}}\right)\,.
\end{equation}
With $k_V=k_R+k_a$, Eqs. \eqref{eq:Btwicorr}--\eqref{eq:ka} fully specify $B_{\rm twi}$.

The sky brightness due to light pollution $B_{\rm lp}$ is taken from a world atlas of artificial night sky brightness \citep{Falchi2016, Falchi2016a}, which is derived from high-resolution satellite data. The values reported in this dataset are the illuminances at the zenith, which is usually the darkest portion of the sky. Although the best-fit function implies that most of the light of a city is emitted in the vertical direction, the average light pollution brightness is likely larger than the reported value. There has been extensive work calculating the horizontal illuminance as a function of the zenith sky brightness, which has found that the horizontal sky brightness is approximately a factor of $\pi$ larger than the zenith sky brightness across a range of sites (see \citealt{Garstang1986, Garstang1989, Kocifaj2015, Falchi2023, Faid2024}). For the sake of definiteness, the light pollution from the world atlas is increased by this factor in these calculations. The light pollution can then be calculated at an arbitrary point on Earth. 

The glare from the Moon {($B_{\rm glare}$)} is constructed from two components---glare in the atmosphere and glare in the eye. In both cases, the glare is produced by light from a bright source scattering as it passes through a medium, either the atmosphere or the aqueous humor of the eye. If $P$ is the phase of the moon (from 0 to 1), then the phase angle is $\alpha=\arccos(2P-1)$ and the brightness of the moon at the top of the atmosphere is (in \unit{\lux}, \citealt{Schaefer1992})
\begin{equation}
    \log I^\star_{\rm Moon}=-0.4\left(0.026\alpha+\num{4e-9}\alpha^4+\num{1.26}\right)\,.
\end{equation}
The brightness of the Moon to an observer is given by 
\begin{equation}
    I_{\rm Moon}=I^\star_{\rm Moon}10^{-0.4k_V(H)X(H, \vec{\theta}_M)}\,,
\end{equation}
where $k_V$ is the extinction {as defined above}, $X$ is the airmass, and $\vec{\theta}_M$ is the direction of the Moon on the sky. Following previous work \citep{Holladay1926, Krisciunas1991, Schaefer1991, Schaefer1992}, the glare of the Moon produced by scattering in the eye (in \unit{cd/m^2}) is 
\begin{equation}
    B_{\rm eye}=\num{13.7}I_{\rm Moon}\theta^{-2}\,,
\end{equation}
where $\theta$ is the separation between the Moon and the {target location in the sky} in degrees. By a similar calculation, the glare {from the Moon due to} scattering in the atmosphere is 
\begin{equation}\label{eq:atmglare}
    B_{\rm atm}=\num{18.5}I_{\rm Moon}^\star\left[10^{-0.4k_VX}-10^{-0.8k_VX}\right]\theta^{-2}\,,
\end{equation}
so long as $\theta\leq\qty{5}{\degree}$. Otherwise, the glare from the atmosphere is set to zero. In Eq. \eqref{eq:atmglare}, $X$ is the airmass between the observer and the target {location} on the sky, not the Moon itself.

Meanwhile, the sky brightness due to the Moon is calculated using an approximate scattering phase function of the atmosphere. The scattering function in the atmosphere is approximately \citep{Krisciunas1991}
\begin{equation}
    f(\theta)\simeq10^{5.36}(1.06+\cos^2\theta)+10^{6.15-(\theta/\qty{40}{\degree})}+\num{6.2e7}\theta^{-2}\,.
\end{equation}
With this approximation, the brightness due to the Moon (in \unit{cd/m^2}) is 
\begin{equation}
    B_{\rm Moon}=f(\theta_{\rm Moon})I_{\rm Moon}\left(1-10^{-0.4k_VX}\right)\,,
\end{equation}
where the final term {in parentheses} accounts for losses in the atmosphere along the scattering path and the airmass $X$ is evaluated at the point of interest $\vec{\theta}$ \citep{Schaefer1993}. If the Moon is below the horizon, then both $B_{\rm Moon}$ and $B_{\rm glare}$ are set to zero.

\begin{figure*}[t]
    \centering
    \includegraphics{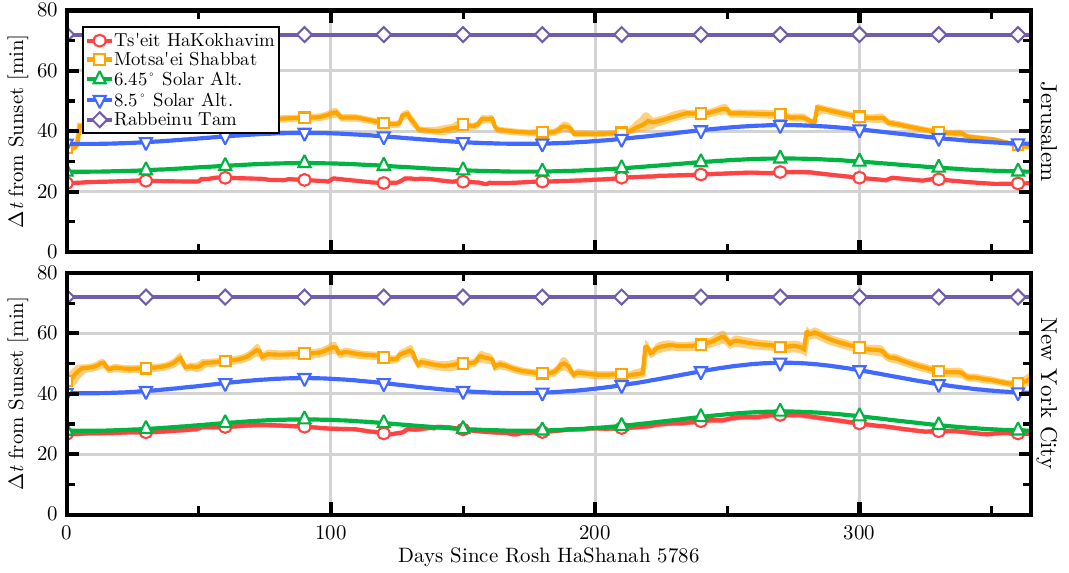}
    \caption{\textbf{Halakhic Times Without Light Pollution.} The halakhic times for Jerusalem {(top panel)} and New York City {(bottom panel)} for the year 5786 {(2025/2026 in the Gregorian calendar)}. This figure shows the difference between {astronomical} sunset and the {calculated} time of \THK (red/circles); \MS (yellow/squares); the time when the Sun is \qty{6.45}{\degree} below the horizon{, which approximates \THK} (green/upwards triangles); the time when the Sun is \qty{8.5}{\degree} below the horizon ({approximating \MS,} blue/downwards triangles); and R. Tam's \qty{72}{\minute} definition (purple/diamonds). The errors in the times of \THK and \MS, assuming a \qty{20}{\percent} error in the sky brightness, are shown as a {lighter} band, but these are generally smaller than the line marking the time. The markers are generally placed at the first of the Jewish month during the new Moon.}
    \label{fig:time_plots}
\end{figure*}

With these definitions in hand, the sky brightness can be calculated at an arbitrary {observer} location, elevation, and time. Similar calculations have an uncertainty in the sky brightness of $\sim\qty{20}{\percent}$, which is assumed to be appropriate for this calculation as well. This is also the approximate accuracy of the sky brightness data at an elevated site. Sky brightness errors of this magnitude {induce errors in the timing of nightfall on the order of a few minutes, which will be discussed later.}

For a given sky brightness $B$ in \unit{\candela\per\meter^2}, the threshold increment illuminance (defined to be the dimmest observable source) $I$ (in \unit{\lux}) is reasonably approximated by (\citealt{Crumey2014}'s empirical fit to the data of \citealt{Blackwell1946})
\begin{equation}
\begin{split}
    I(B)=\big(&\sqrt{a_1B^{1/2}+a_2B^{3/4}+a_3B}\\
    &+a_4B^{1/4}+a_5B^{1/2}\big)^2\,. \label{eq:minI}
\end{split}
\end{equation}
The empirical constants are 
\begin{equation}
    \begin{split}
        &a_1=\num{6.112e-8};\,a_2=\num{-1.598e-7};\\
        &a_3=\num{1.167e-7};\,a_4=\num{4.988e-4};\\
        &a_5=\num{-3.014e-4}\,.
    \end{split}
\end{equation}
The star brightness necessary for observability $I$ is related to the requisite brightness at top of the atmosphere $I^\star$ by 
\begin{equation}
    I^\star=I\,10^{0.4k_V(H)X(H,\vec{\theta})}\,.
\end{equation}
The corresponding limiting magnitude is then \citep{Cox2002}
\begin{equation}\label{eq:maglim}
    m=-2.5\log\Delta I-13.99\,.
\end{equation}
This magnitude is reduced by one, since the observer is likely using directed vision to find small or medium stars.

The Yale Bright Star Catalog \citep{Hoffleit1991} is used to provide a list of the positions and magnitudes of all \num{126} medium ($\qty{3}{mag}\leq m<\qty{2}{mag}$) and \num{4910} small ($\qty{6}{mag}\leq m<\qty{3}{mag}$) stars. For a given observer location, elevation, and time, the location of each star on the sky is calculated using the \texttt{PyEphem} package \citep{Rhodes2011}. Each star greater than \qty{10}{\degree} above the horizon and with a magnitude less than {(brighter than)} the limiting noticeable magnitude is counted as visible. If there are at least three visible medium stars, then \THK has already passed. If there are at least three small stars that are (i) visible and (ii) within \qty{10}{\degree} of each other (satisfying the ``in one place'' condition imposed by Shulchan Aruch Orach Chaim 293:2), then \MS has already passed. For a given date, the precise moment when these times occur is found by {calculating if these conditions have been met} in increments of \qty{10}{min} {after sunset}. Once the {relevant conditions are met}, a binary search is performed in the final \qty{10}{min} range to calculate the time to within \qty{1}{s}. This advance-search method will always find the first {instance} that the nightfall conditions are met. This algorithm accounts for cases where three stars are not visible later at night due to the setting of stars or rising of the Moon.

An implementation of this algorithm is available at \href{halakhic-nightfall.streamlit.app}{halakhic-nightfall.streamlit.app}. This tool can calculate the times of \THK and \MS at an arbitrary location on Earth on an arbitrary date. If unspecified, the elevation at the input location is calculated from the {Mapzen open-source elevation database}. \texttt{PyEphem} is also used to calculate the time of sunset and the two relevant solar altitude approximations. 

\begin{figure*}
    \centering
    \includegraphics{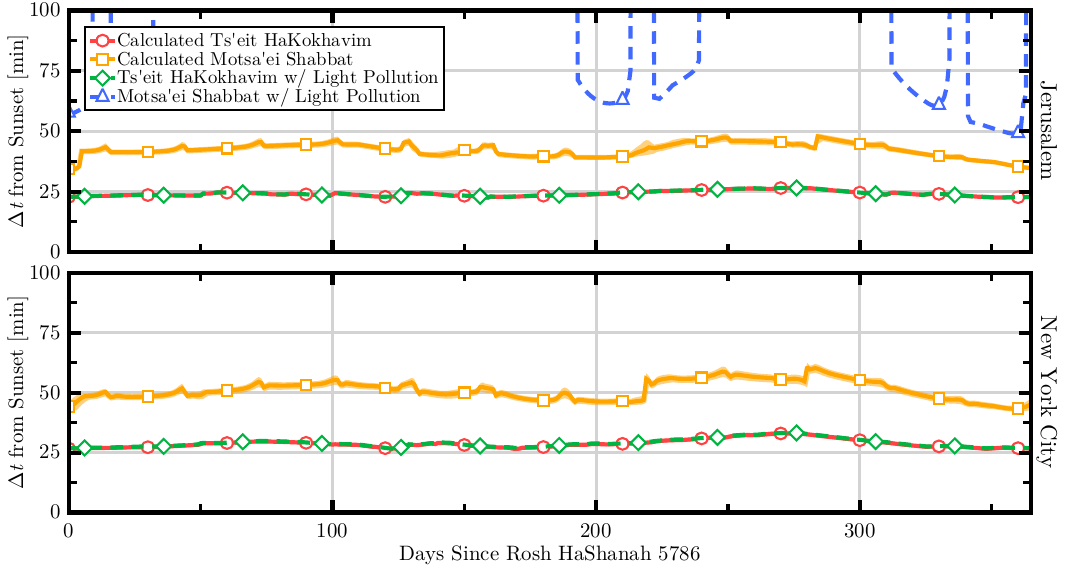}
    \caption{\textbf{Effect of Light Pollution on Halakhic Times.} The halakhic times for Jerusalem {(top panel)} and New York City {(bottom panel)} for the year 5786, accounting for light pollution. This figure shows the difference between {astronomical} sunset (zero on the $y$-axis) and \THK (red/circles), \MS (yellow/squares), \THK when light pollution is included (green/diamonds), and \MS when light pollution is included (blue/triangles). {The} errors in the times of \THK and \MS, assuming a \qty{20}{\percent} error in the sky brightness, are shown as a {lighter} band, which is generally obscured by the plotted line. Light pollution has an effect on the time of \THK of approximately six seconds. In New York City, {the conditions for} \MS never occur when light pollution is included{, and only occasionally occur in Jerusalem.}}
    \label{fig:lp_plots}
\end{figure*}

\section{Timing Comparison}\label{sec:tcomp}

This tool is used to calculate the nightfall time every night for the Jewish year 5786, which is 2025/2026 in the Gregorian calendar. Fig. \ref{fig:time_plots} shows the difference between sunset and the various halakhic times (without light pollution) for two locations --- Jerusalem and New York City --- over the course of the year. In addition to the times of \THK and \MS, this figure shows the \qty{6.45}{\degree} solar altitude approximation for \THK, the \qty{8.5}{\degree} solar altitude approximation for \MS, and R. Tam's \qty{72}{\minute} approximation. 
Throughout the remainder of this text, ``\THK'' and ``\MS'' will refer to the times calculated by the algorithm presented above. 
This phrasing should not be interpreted as a halakhic ruling on the ``true'' conditions, which is well beyond the scope of this work.

Several conclusions can be drawn from this figure. 
First, the time of \THK is very similar to the \qty{6.45}{\degree} solar altitude approximation, although the time of \THK has small jumps from night to night that are not seen in the approximation time. 
A solar altitude approximation could be exactly correct if (i) the same stars are in the sky from night to night and (ii) that the third-brightest star is in the same place each night. 
Recall that \THK occurs when any three medium stars are visible irrespective of the stars' positions. 
Since there is a direct and monotonic relationship between the sky brightness and the visibility of stars, the brightest star in the sky will be visible first, then the second, then the third. 
Therefore, the third-brightest medium star in the sky will always be the critical point for \THK. 
\THK will thus occur when the sky's limiting magnitude is equal to the magnitude of the third-brightest star. 
Since the sky brightness at a site depends almost entirely on the Sun's altitude, so long as this star is in the same place in the sky from night to night, then \THK will always occur when the Sun is at a specific altitude. 
In this case, there would exist an exactly correct solar altitude approximation. 

The small differences between the time of \THK and a solar altitude approximation thus arise because of changes in the assumptions underlying the above logic chain. 
As the Earth moves relative to the Sun over the course of the year, the position of the stars shifts with respect to the Sun. 
Stars that are in the sky at midnight in the winter will be in the sky at noon in the summer. 
Therefore, the three brightest stars in the sky will not necessarily be identical from night to night. 
Most critical is the position of the third-brightest star, since this sets the actual time of \THK. 
As the third-brightest star in the sky at sunset changes, different stars will set the time of \THK. 
Each star will have a slightly different brightness and therefore require slightly different sky brightnesses and solar altitudes to be visible. 
However, since the brightest medium stars all have similar magnitudes, a solar altitude approximation is reasonably accurate. 

\begin{figure*}
    \centering
    \includegraphics{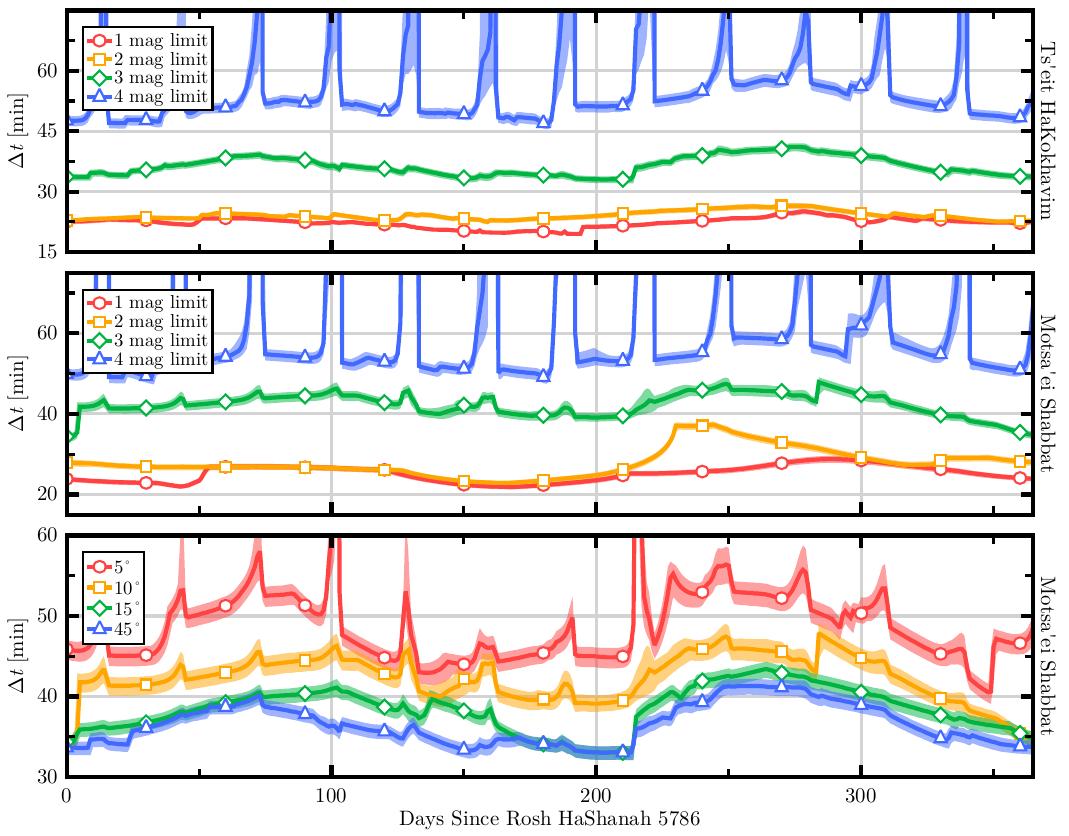}
    \caption{\textbf{Effects of Varying Parameters on Nightfall Timing.} The {timing of \THK and \MS} in Jerusalem for the year 5786 using different definitions of the brightest relevant star and of ``in one place''. {Light pollution is not included.} The top panel shows the variation in the timing of \THK if the upper limit for medium stars is varied from \qtyrange{1}{4}{mag}. In this work, the time of \THK uses a \qty{2}{mag} cutoff by default. {The middle panel shows the time of \MS if the upper limit for small stars is varied over the same range, with the default being \qty{3}{mag} for \MS. Finally, the bottom panel shows the timing of \MS if the maximum distance at which stars are assumed to be ``in one place'' is changed from the default \qty{10}{\degree}.} The variance from assuming a \qty{20}{\percent} error in the sky brightness is shown as a {lighter} band. Note the differences in the scale of the $y$-axes.}
    \label{fig:parcomp}
\end{figure*}

On the other hand, while the time of \MS is only about \qty{5}{\minute} later than the \qty{8.5}{\degree} solar altitude approximation, the solar altitude approximation {curve has a fundamentally different shape than the \MS curve}. These more significant differences result from the fact that (i) these {small} stars are dimmer and thus are more sensitive to small changes in the sky brightness and (ii) the limitation that the three stars are ``in one place''. Since the stars are {smaller}, dimmer skies are necessary for these stars to be visible. The brightness of the Moon is much more significant for these skies, so the timing of \MS is much more sensitive to the phase of the Moon. The small regular spikes in the timing of \MS occur at the full Moon {(shown best in Fig. \ref{fig:parcomp})}, when the night sky is significantly brighter.

The impact of the ``in one place'' condition is somewhat more complex. Consider all small stars that will be in the sky on a given night. These stars can be grouped into sets of three that are within \qty{10}{\degree} of each other and which would satisfy the ``in one place'' condition. Any star not in one of these {triads} is not relevant to calculating \MS. \MS will occur when the sky is dim enough that all three stars in a {specific triad} are visible. Note that there may be significantly more than three stars visible, since it is very unlikely that the three brightest {small} stars in the sky are within \qty{10}{\degree} of each other. Since the limiting magnitude is monotonic in sky brightness, all stars in a {triad} will be visible once the dimmest star in that set can be seen. Therefore, \MS will occur when the brightest star that is the dimmest star in its {triad} has become visible. 

Since the stars in the sky are not the same from night to night, \MS will not occur at a fixed solar altitude. 
In contrast to the {timing of} \THK, the shifting of the stars is far more significant. 
For \THK, the set of ``critical stars'' (the stars that, by becoming visible, satisfy the conditions for nightfall) is the set of all stars that are, on some night, the third brightest medium star in the sky. 
These stars will have some variation in brightness, but will be largely similar. 
On the other hand, the critical stars for \MS compose the set of the brightest small stars that are the dimmest in their triad. 
Since \MS has an additional condition that is independent of the brightness, these stars will represent a broader range of brightnesses. 
Over the course of the year, the sky brightness when \MS occurs will therefore vary significantly as different critical sets are used. 

The standard \qty{8.5}{\degree} solar altitude approximation thus provides a poor estimate for the time of \MS from night to night. 
In addition, \MS generally occurs after the Sun reaches an altitude of \qty{-8.5}{\degree}. Using this approximation therefore underestimates the length of Shabbat and has serious implications for proscribed actions. 
{While} R. Tam's approximation is always later than \MS and provides a safe alternate estimate, its suitability beyond its relation to the calculated times is outside the scope of this paper. 

\begin{table}[t]
    \centering
    \caption{\textbf{Effect of Light Pollution.} The maximum change in the time of \THK induced by light pollution in the year 5786 (2025/2026) at several {locations}. In all listed locations, light pollution causes \MS to not occur on at least some days {in the year}. {Therefore, the listed time is only the effect on \THK (\textit{T'H}).}}
    \begin{tabular}{c|rrr}
        City & Lat. & Long. & $\Delta t$ {(\textit{T'H})} \\\hline
        New York & \qty{40.7128}{\degree N} & \qty{74.0060}{\degree W} & \qty{0.47}{\minute}\\
        Buenos Aires & \qty{34.6037}{\degree S} & \qty{58.3816}{\degree W} & \qty{0.47}{\minute}\\
        Los Angeles & \qty{34.0549}{\degree N} & \qty{118.2426}{\degree W} & \qty{0.27}{\minute}\\
        Reykjavik & \qty{6.1470}{\degree N} & \qty{21.9408}{\degree W} & \qty{0.27}{\minute} \\
        Paris & \qty{48.8575}{\degree N} & \qty{2.3514}{\degree E} & \qty{0.27}{\minute}\\
        London & \qty{51.5074}{\degree N} & \qty{0.1278}{\degree W} & \qty{0.27}{\minute}\\
        Jerusalem & \qty{31.7769}{\degree N} & \qty{35.2345}{\degree E} & \qty{0.20}{\minute}\\
        Alexandria & \qty{31.2001}{\degree N} & \qty{29.9187}{\degree E} & \qty{0.12}{\minute}\\
        Ann Arbor & \qty{42.2808}{\degree N} & \qty{83.7430}{\degree W} & \qty{0.08}{\minute}\\
    \end{tabular}
    \label{tab:lp_change}
\end{table}

\section{Light Pollution}\label{sec:lp}

Over the last century and a half, the profusion of electric lighting has lead to increasing light pollution across the Earth's surface. 
Today, only \qty{20}{\percent} of the {global population} and only \qty{1}{\percent} of Europeans and Americans live under pristine skies \citep{Falchi2016a}. 

In addition to the previously calculated timings of \THK and \MS, Fig. \ref{fig:lp_plots} shows the nightfall times {in Jerusalem and New York City} when light pollution is accounted for. Table \ref{tab:lp_change} presents the largest delay in \THK produced by light pollution during the year 5786 for several sites. 

In general, light pollution has an effect on the timing of \THK on the order of less than a minute. 
After all, even in the most light polluted sites, stars of magnitude $m\geq\qty{3}{mag}$ are readily visible. 
Since this is the dimmest possible medium star, light pollution will only slightly delay the time of \THK. 
{Note that since light pollution is added to the sky brightness, it can only delay the timing of \THK and \MS, not hasten their arrival.}

In contrast to its effects on \THK, light pollution has a significant impact on the observation of \MS. Since \MS requires three small stars to be observable and close together, the necessary conditions are usually only met when stars of approximately \qty{4}{mag} become visible. 
In Jerusalem, light pollution is significant enough that critical star is required to be relatively bright, causing the conditions to only be met occasionally throughout the year. 
However, the more significant light pollution in New York City means that the conditions for \MS never occur in 5786.
{In all locations studied in Table \ref{tab:lp_change}, there are at least some days when light pollution causes the conditions for \MS to never be met throughout the night.
This pattern will likely continue in future years.
Depending on how \MS is halakhically defined, this may have significant impacts on Jewish practice.}

\section{Definition Changes}\label{sec:pareffs}

While this work argues that a ``medium'' star is \qtyrange{2}{3}{mag} and that stars closer than \qty{10}{\degree} are ``in one place'', these definitions are subject to debate. This section thus discusses the impact that varying these parameters has on the timing of \THK and \MS. 

Fig. \ref{fig:parcomp} shows the time of \THK and \MS in Jerusalem (ignoring light pollution) for a range of parameters. 
The top panel shows the variations that occur in the {timing of \THK when the maximum brightness of a medium star is changed. The middle panel shows the behavior of \MS when the size of small stars is changed, while the bottom panel shows how \MS is affected by different definitions of the ``in one place'' condition.} When changing the magnitude, the dimmest star considered was always set to two magnitudes less than the brightest star. Since the brightest stars are visible first, changing the cutoff of the dimmest stars has no impact as long as there are a reasonable number of stars within the defined range. 

Increasing the magnitude cutoff (so that brighter stars are ignored) delays the time when three stars are visible. Since there are only \num{35} stars between \qty{1}{mag} and \qty{2}{mag}, using one or the other generally has little difference in the timing of nightfall. However, setting the cutoff to \qty{3}{mag} or \qty{4}{mag} has significant effects for both \THK and \MS. {This is clearly visible in the top two panels of Fig. \ref{fig:parcomp}.}

As the brightest allowed star gets dimmer, the impact of the phase of the Moon gains significance. If the brightest star is \qty{4}{mag}, then three stars are never visible {in Jerusalem} during a full Moon. After all, a full Moon sets the limiting noticeable magnitude in much of the sky to nearly \qty{4}{mag}, which makes it impossible to see three stars dimmer than this cutoff. Even for a cutoff of \qty{3}{mag} (used for \MS), three stars are visible earlier during a new Moon than a full Moon. 

Finally, consider the variations that occur when the ``in one place'' definition is changed. As expected, increasing the cutoff distance will make three stars visible earlier. After all, any set of three stars that are within a \qty{10}{\degree} radius circle are also within a \qty{15}{\degree} radius circle, and so on. 
As the distance cutoff continues to increase, the time of \MS approaches the value that it would have in the absence of the ``in one place'' restriction altogether (as is the case for \THK). {This can be seen by comparing the blue curve (triangle markers) in the bottom panel of Fig. \ref{fig:parcomp} with the green curve (diamond markers) in the top panel. Since \qty{45}{\degree} is a large cutoff distance and both of these curves use a \qty{3}{mag} upper limit, they are nearly identical. Note that the visual differences in these curves are caused by the different axis scales between panels. Finally, since \qty{5}{\degree} is a relatively small cutoff, on some days the only triads that satisfy the distance conditions consist of relatively dim stars. In this case, the full Moon can render these triads unobservable, and \MS will never occur, even in the absence of light pollution. }

\section{Conclusion}\label{sec:conc}

\subsection{Summary}\label{sec:sum}

This work has used modern astronomical techniques to investigate the problem of {accurately} calculating \THK and \MS. A tool for calculating these times is available online at \href{halakhic-nightfall.streamlit.app}{halakhic-nightfall.streamlit.app}. I have compared these calculated times to the approximations most commonly used by the Jewish community, and found that the standard solar altitude approximation generally predicts that \THK is later than when the conditions are actually met by approximately five minutes. However, the \qty{8.5}{\degree} solar altitude approximation underestimates the length of Shabbat and the time of \MS {(again by five minutes)}, which can create difficulties for the observant Jew.

This tool was also used to calculate the effects of light pollution on the observation of \THK and \MS. Light pollution has little effect on the timing of \THK{, on the order of a minute at most}. After all, some stars are visible even in the brightest cities. On the other hand, light pollution has a significant impact on the time of \MS. The conditions for \MS rarely occur in Jerusalem and never in New York City if light pollution is accounted for. 
{In all cities considered, there is at least one day a year in which light pollution means that the conditions for \MS are never met.}
{Note as well that since a medium star is defined to be \qtyrange{2}{3}{mag} and a small star is \qtyrange{3}{6}{mag}, is difficult to define a cutoff for small stars that ensures that small stars are always visible without encroaching on medium stars.}

\subsection{Discussion}\label{sec:disc}

While this tool provides an improved calculation for the time of \THK and \MS, there remains some uncertainty in {the calculated times}. First, the background sky brightness {is} a function of the air quality, cloud cover, the color of the stars \citep{Bara2020}, and small-scale light pollution such as streetlights \citep{Bara2023}, which can impact the calculation of visibility. Fortunately, many of these unaccounted-for factors produce changes of less than the estimated \qty{20}{\percent} uncertainty in the sky brightness and will thus have little impact on {calculations of} the timing of nightfall. Second, the observer may know the location of the three most relevant stars, allowing them to precisely search the sky and observe stars earlier than an uninformed searcher. 

It is also necessary to question the utility of \MS's distinction. \MS is only different from \THK because of the additional stringencies imposed by the Shulchan Aruch (Orach Chaim 293:2). The Mishna Brurah (293:2) explains that this stringency is in place because nobody has the expertise to make the distinction between medium and small stars. If one accepts the {definitions of star size and ``in one place''} made in Sec. \ref{sec:condspec}, then {it can be argued that} this additional stringency is no longer relevant. {Any layman using this calculation will obtain the same results as an expert who is using these definitions, meaning that individual expertise is no longer necessary. 
Alternatively, anyone who understands the magnitude system and the algorithm behind these calculations could be considered to have the expertise necessary to calculate nightfall when using this tool.
If this argument holds, then the end of Shabbat can be said to occur at \THK.} Of course, {much} discussion must be had before such a significant change is made to the halakhah.

A more critical factor is the uncertainty in the definitions of star sizes and ``in one place''. This calculation assumes that a star between magnitude 2 and 3 is ``medium'' and that a star less than magnitude 3 is ``small''. Note that the naked-eye limiting magnitude is 6 in the darkest skies, and the fifth-brightest star in the sky is Vega, at magnitude 0. While these definitions are {derived from} the text of the Talmud, I encourage additional discussion on the appropriate definition. In addition, the ``in one place'' condition is interpreted here to mean that the three relevant stars must be within \qty{10}{\degree} of one another. This number is chosen to represent a region of the sky where all of the stars can be held in the foveal field without moving one's eyes. While other possible distance values can be used --- which will change the timing of \MS --- a solar altitude approximation will never be astronomically accurate due to the movement of the stars over the course of a year. 

If there is disagreement on the precise halakhic definitions, the software used to calculate the timing of nightfall can be easily modified to accommodate different limits on brightness and distance. {Even if there is disagreement on the precise magnitude cutoff between a medium and small star or on the distance constraint implied by ``in one place'', I believe that the introduction of this framework significantly aids in the relevant discussions.} {The use of magnitudes and degrees} provides a concrete description of this previously-vague issue and enables a more rigorous discussion of the timing of \THK and \MS.

{The fact that light pollution causes the conditions for \MS to never occur on some days has significant implications for Jewish practice. If one is required to be able to observe three stars regardless of light pollution before Shabbat ends, then Shabbat will extend throughout much of the year in most locations, and never end in very light-polluted cities like New York. Such a scenario is untenable, and halakhic authorities must address this issue. While there are many possible approaches, a few are discussed below.}

First, one could assume that the Shabbat stringencies never end, and follow the Shabbat proscriptions throughout the week. This would require most observant Jews to quit their jobs, refrain from cooking, and spend their time studying Torah. This is unrealistic for most Jews, would lessen the sanctity of the Shabbat, and should be avoided.

Second, one could rule that the long tradition of the solar altitude approximations means that they have become the true definitions of \THK and \MS, supplanting the description of the Talmud. In this case, the approximations would of course function perfectly and Shabbat will end at the calculated time. While this is a tempting proposition, it must be noted that the \qty{8.5}{\degree} solar altitude approximation generally predicts that Shabbat ends approximately five to ten minutes before the conditions of \MS are met, even in the absence of light pollution. While small, this difference may matter to some communities or individuals.

A third path could be to rule that light pollution should be disregarded when calculating \MS. This approach is justified by the Shulchan Aruch's ruling for overcast days --- namely, that one should wait until they are no longer uncertain that three stars have appeared (Orach Chaim 293:2). The star-obscuring effects of light pollution could be considered to have similar properties to those of clouds, so that \MS would occur when three stars would be visible in the absence of light pollution. The time of \THK and \MS could therefore be calculated directly by \href{halakhic-nightfall.streamlit.app}{halakhic-nightfall.streamlit.app} or a similar tool, which ignores clouds and provides results with and without light pollution. The errors in the sky brightness could be safely accounted for by adding ten minutes to the calculated times.

This issue is particularly salient given the fact that no city could be found where the conditions for \MS (when including light pollution) occur in every day of 5786. While \MS is only relevant one day a week, light pollution generally means that \MS will fail to occur for more than a week. 
At the latitude of Jerusalem, a background light pollution brightness of $B\sim\qty{0.01}{cd/m^2}$ is necessary for \MS's conditions to occur every day. 
This limit corresponds to a suburban sky in the Bortle scale, so that even suburban areas will likely fail to meet the conditions for \MS. Even in suburban areas where these conditions are met every day, light pollution will often delay it by many hours. 
Note that since Shabbat restrictions begin at sunset (not \THK or \MS), Shabbat is lengthened, not shifted. 
Only Jews in rural environments will regularly experience \MS near when it would occur without light pollution. 
However, since Jewish communities are generally urban or suburban, most of the Jewish population will experience at least one day a year where the conditions for \MS does not occur and will regularly experience a significant delay before the conditions for \MS are met. 

\section*{Acknowledgments}

I thank Leo Barry for his invaluable support, insightful comments, and great help in improving the clarity and direction of this paper. Shannon Murphy very kindly lent me her sky quality meter for the measurements. I also thank Drew Weisserman, Liyam Chitayat, and Darryl Seligman for helpful conversations. I acknowledge support from the Fannie and John Hertz Foundation and the University of Michigan's Rackham Merit Fellowship Program.

\begin{figure}[h]
    \centering
    \includegraphics{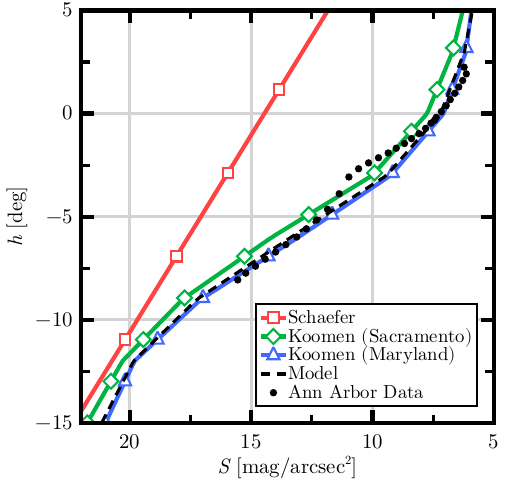}
    \caption{\textbf{Zenith Brightness Comparison.} The zenith brightness versus solar altitude using the data of \citet{Koomen1952} (green diamonds and blue triangles for the Sacramento Peak and Maryland sites, respectively), the empirical formula of \citet{Schaefer1987} (red squares), and novel measurements of the sky brightness in Ann Arbor, MI (black). The uncertainty in the measurements is smaller than the shown data points. }
    \label{fig:bcomp}
\end{figure}

\appendix

\section{Twilight Brightness Measurements}\label{sec:twibrimeas}

This section presents a comparison between the empirical twilight brightness data of \citet{Koomen1952}, the formula of \citet{Schaefer1987}, and empirical data that was obtained by the author. These results clearly indicate that a fit to empirical brightness data is a far superior match to the measurements than the empirically-calibrated formula used in previous sky brightness calculators. 

In the \citet{Schaefer1987} approximation, the sky brightness $S$ in \unit{mag/arcsec^2} is given by 
\begin{equation}
    S = 14.45+0.83\theta Z-h(30+20.53Z)-7.15Z\,,
\end{equation}
where $\theta$ is the azimuth difference between the Sun and the pointing location, $Z$ is the zenith distance, and $h$ is the Sun's altitude (all in radians). At the zenith, $Z=0$ and $\theta$ is unspecified so that $S=14.45-30h$. 

Measurements of the zenith sky brightness were taken using a UniHedron Sky Quality Meter, which measures the brightness of the sky within a range of \qty{60}{\degree} of the zenith. Measurements were recorded every minute to ensure a consistent sample, with the Sun's altitude calculated using \texttt{PyEphem}. The uncertainty in each measurement is $\pm\qty{10}{\percent}$, or $\pm\qty{0.10}{mag/arcsec^2}$. Data were taken on several clear nights in Ann Arbor, MI (elevation \qty{262}{m}). The Koomen data also depend on elevation, with the Maryland site at \qty{30}{m} and the Sacramento Peak, NM site at \qty{2800}{m}. The elevation is accounted for using the method described in Sec. \ref{sec:skybri}. 

The new measurements, the Koomen data, and the Schaefer model are compared in Fig. \ref{fig:bcomp}. In addition to the data, models of the sky brightness (calculated from the Koomen data) at the correct elevation of the novel {Ann Arbor} measurements are shown as a thin dashed line. The uncertainty in the measured sky brightness is plotted, but is smaller than the data points. Although there are some features of the brightness curve that are produced by local geography and not captured by the model,\footnote{Due to practical constraints, these measurements were not made at local geographic high points.} the Koomen data is clearly a superior fit to the sky brightness than the Schaefer model {for the purposes of this work}. The Schaefer model will, in fact, consistently predict a significantly dimmer zenith sky than the reality and imply that far more stars are visible at sunset than is possible. Of course, the Schaefer model is specifically calibrated to reproduce the sky brightness near the Western horizon during twilight, and so is not intended to reproduce the zenith brightness. Despite its intended purpose, the Schaefer model is used to calculate the brightness of the entire twilight sky in many commercial and open-source calculators. To avoid this issue, {an} interpolation of the Koomen data is used throughout the text.

\nocite{ShulchanAruch}
\nocite{BiurHaGra}
\nocite{SeferIgrotMoshe}
\nocite{SeferBenHaShemashot}
\nocite{PHShabbat}

\bibliography{main}{}
\bibliographystyle{chicago}
\end{document}